\documentclass[letterpaper,english,reprint, aps]{revtex4-1}
\usepackage[T1]{fontenc}
\usepackage[latin9]{inputenc}
\usepackage{textcomp}
\usepackage{mathrsfs}
\usepackage{amsmath}
\usepackage{amssymb}
\usepackage{graphicx}

\makeatletter

\pdfpageheight\paperheight
\pdfpagewidth\paperwidth

\makeatother

\usepackage{babel}
\begin{document}

\title{Nonlocal black hole evaporation and quantum metric fluctuations via
inhomogeneous vacuum density}

\author{Alexander Y. Yosifov}
\email{alexanderyyosifov@gmail.com}

\affiliation{Department of Physics and Astronomy, Shumen University}

\author{Lachezar G. Filipov}
\email{lfilipov@mail.space.bas.bg}

\affiliation{Space Research and Technology Institute, Bulgarian Academy of Sciences}
\begin{abstract}
Inhomogeneity of the \emph{actual} value of the vacuum energy density
is considered in a black hole background. We examine the back-reaction
of a Schwarzschild black hole to the highly inhomogeneous vacuum density,
and argue the fluctuations lead to deviations from general relativity
in the near-horizon region. In particular, we found that vacuum fluctuations
\emph{onto} the horizon trigger adiabatic release of quantum information,
while vacuum fluctuations in the vicinity of the horizon produce potentially
observable metric fluctuations of order the Schwarzschild radius.
Consequently, we propose a form of strong nonviolent nonlocality in
which we simultaneously get nonlocal release of quantum information,
and observable metric fluctuations. 
\begin{description}
\item [{PACS~numbers:}] 04.70.Dy, 04.60.\textminus m{\small \par}
\item [{Keywords:}] black holes, quantum gravity, gravitational waves{\small \par}
\end{description}
\end{abstract}

\pacs{04.70.Dy, 04.60.\textminus m}

\keywords{black holes, quantum gravity, gravitational waves}
\maketitle

\subsection*{1. Introduction}

Recently, Unruh et al. argued (see Ref. {[}1{]}) the observed small
non-negative cosmological constant can be achieved without introducing
new degrees of freedom, \emph{e.g.} negative-pressure scalar field.
Instead, they address the cosmological constant problem by embracing
the diverging value of the vacuum energy density as predicted by quantum
field theory, without applying any renormalization procedures. Interestingly,
by studying the gravitational effects of the constantly fluctuating
vacuum, they found its local energy density to be highly inhomogeneous.
As a result of this inhomogeneity, the spatial distance between a
pair of neighboring points undergoes constant phase transitions in
the form of rapid changes between expansion and contraction. Also,
the singular expectation value of the local energy density of the
vacuum was argued to be harmless at the energy levels of effective
field theory since the fluctuations lead to huge cancellations on
cosmological scales, and ultimately to the observed accelerating expansion
of the universe. 

In a separate work {[}2{]}, we studied the back-reaction of a Schwarzschild
black hole geometry to quantum vacuum fluctuations. We examined how
the black hole metric back-reacts to vacuum fluctuations in two regions,
\emph{onto} the horizon, and in the vicinity of the black hole. We
considered vacuum fluctuations with local energy density \emph{below},
and \emph{above} a certain threshold $\zeta$. We found that \emph{\textquotedbl{}strong\textquotedbl{}}
quantum fluctuations (considered onto the horizon) lead to brief nonviolent
departures from classicality, which allow for adiabatic leakage of
low-temperature Hawking quanta at the necessary rate. In addition,
we argued that quantum information can begin leaking out of the black
hole as early as $r_{S}\log r_{S}$ after the initial collapse. The
\emph{\textquotedbl{}weak\textquotedbl{}} quantum fluctuations (considered
in the vicinity of the black hole) were argued to be the microscopic
source, which on scales of order the Schwarzschild radius $\mathcal{O}(r_{S})$
accumulates and produces metric fluctuations {[}2,3{]}. 

The current model questions the classical black hole picture of local
quantum field theory on a semiclassical geometry. In this work we
apply the Unruh et al. model to Schwarzschild geometry; namely, we
rewrite the equations of {[}1{]} in that background, and study the
gravitational effects of the inhomogeneous vacuum density. As a result,
we propose a form of strong nonviolent nonlocality which yields significant
modifications to the well-known general relativistic picture of black
holes. The conjectured deviations from classicality lead simultaneously
to nonlocal release of Hawking particles, and quantum metric fluctuations.
In fact, the present work may be thought of as the microscopic origin
of the initially proposed by Giddings nonviolent nonlocality model
{[}4,5,6{]}.

The paper is organized as follows. In Section 2 we summarize the relevant
work. In Section 3 we rewrite the equations of {[}1{]} in Schwarzschild
black hole background, and show that we can derive non-trivial modifications
to general relativity in the near-horizon region.

\subsection*{2. Summary of related work}

To make the paper self-contained, we begin by making a brief review
of the main results of {[}1{]} and {[}2{]}. Note the Unruh et al.
model will serve as a basis for the scenario we will present later
in Section 3. In Section 2.1 we demonstrate the physical interpretation
of the conjectured in {[}1{]} inhomogeneous microscopically diverging
vacuum energy density, and its effects on cosmological scales. Then,
in Section 2.2, we review {[}2{]} to show how we initially derived
the \textquotedbl{}soft\textquotedbl{} but non-trivial modifications
to general relativity in the near-horizon region.

\subsection*{2.1. Accelerating expansion via inhomogeneous quantum vacuum density}

In the generic $\varLambda$CDM model of the universe the vacuum energy
density is considered to be constant and homogeneous throughout space.
As it was pointed out in {[}1{]}, however, this basic assumption is
only true for the expectation value of the vacuum energy density.
Its \emph{actual} value, \emph{i.e}. the one obtained by performing
repeated measurements at a particular spatial point, constantly fluctuates.
Consequently, we get a picture, where although the expectation value
is effectively constant and homogeneous on cosmological scales, the
actual value is rapidly changing in both time, as well as from point
to point. Physically, this inhomogeneity of the vacuum density implies
the spatial distance between any pair of nearby points constantly
changes between phases of expansion and contraction. As we show in
greater detail later, this leads to very important results in a black
hole background.

Taking the constantly fluctuating inhomogeneous vacuum density as
a starting point, we now briefly demonstrate its effects in a general
spacetime {[}1{]}. Note that in Section 3 we apply those results to
Schwarzschild black hole geometry.

Suppose the local energy density $\rho_{x,x'}$ between a pair of
neighboring spatial points $x$ and $x'$ in some general metric $g_{\mu\nu}$
is {[}1{]}

\begin{equation}
\rho_{x,x'}=\frac{cov(T_{00}(x),T_{00}(x')}{\sigma_{x}\sigma_{x'}}
\end{equation}
where

\begin{equation}
\sigma_{x}=\sqrt{\left\langle (T_{00}(x)-\left\langle T_{00}(x)\right\rangle )^{2}\right\rangle }
\end{equation}
Here, $T_{00}(x)$ and $T_{00}(x')$ are the local vacuum densities
defined at the spatial points $x$ and $x'$, respectively. 

Evidently from (1), the value of $\rho_{x,x'}$ is determined by the
covariant vacuum densities, defined at the neighboring spatial coordinates,
namely $T_{00}(x)$ and $T_{00}(x')$. Effectively, one can think
of $\rho_{x,x'}$ as a 2-point correlation function. That is, in order
for $\rho_{x,x'}$ to have a non-trivial value, it must always be
evaluated between close spacetime points. Otherwise, $\rho_{x,x'}\rightarrow0$
as the separation between $x$ and $x'$ becomes large, in which case
$T_{00}(x)$ and $T_{00}(x')$ are no longer correlated, and thus
evolve independently. The requirement that $x$ and $x'$ be close
comes from the limited domain of dependence of individual vacuum fluctuations;
that is, their high momentum/short wavelength. 

Assuming $\rho_{x,x'}$ is non-vanishing, then in order for it to
be positive/negative, both $T_{00}(x)$ and $T_{00}(x')$ need to
be, respectively, above/below the zero threshold of $\left\langle T_{00}\right\rangle $
{[}1{]}

\begin{equation}
\rho_{x,x'}=\begin{cases}
>0\quad if & T_{00}(x),T_{00}(x')\,>0\\
<0\quad if & T_{00}(x),T_{00}(x')\,<0
\end{cases}
\end{equation}
Therefore, the coefficient $\rho_{x,x'}$ shows the correlation between
vacuum densities defined between a pair of nearby points $x$ and
$x'$. One should note that since in this model we do not apply any
renormalization procedures, we cannot use the generic stress-energy
tensor as a source in the Einstein field equations. Instead, we must
slightly modify the stress tensor in order to account for the diverging
expectation value of the vacuum fluctuations. 

Studying the gravitational effects of the inhomogeneous vacuum density
requires inhomogeneity of the underlying metric as well. So in this
scenario, the scale factor has to have an extra stochastic component
which would allow it to account for that inhomogeneity. Therefore,
following (1), the generic scale factor of the standard Friedmann-Robertson-Walker
metric is modified as

\begin{equation}
ds^{2}=-dt^{2}+a^{2}(t,x)(dx^{2}+dy^{2}+dz^{2})
\end{equation}
As a result of the vacuum inhomogeneity, the scale factor $a(t,x)$
now has additional degrees of freedom in the form of a space-dependent
coupling term. That is, when the local scale factor is evaluated at
a given spacetime point, its dynamics is dictated (sourced) by the
stochastically varying vacuum fluctuations at that point. This richer
structure of the scale factor allows for a pair of nearby points to
be expanding or contracting, depending on the sign of $\rho_{x,x'}$,
(3). In particular, for $\rho_{x,x'}>0$ the spatial separation between
the pair of points increases, while for $\rho_{x,x'}<0$, the spatial
separation decreases.

It was then shown in {[}1{]} that one could consider a local Hubble
rate term, and evaluate it between the neighboring $x$ and $x'$ 

\begin{equation}
\nabla H=-4\pi GJ
\end{equation}
where $H\equiv\dot{a}/a$ is the Hubble parameter, and $J$ denotes
the energy flux of the vacuum, which accumulates over a given region
of space. Hence $J$ can be thought of as a functional of the local
energy density in that neighborhood $\left\langle J(x,x')\right\rangle \sim\int_{x}^{x'}\rho_{x,x'}$. 

Interestingly enough, following the extra degrees of freedom of $a(t,x)$,
$\nabla H$ was found to be \emph{constantly} fluctuating with energy,
sourced by the accumulation of the vacuum density.

The general solution of (5) reads

\begin{equation}
H(t,x)=H(t,x_{0})-4\pi G\int_{x}^{x'}J(t,x')(dx',dy',dz')
\end{equation}
As (6) shows, the local Hubble rate depends on the spatial accumulation
of $J$ in the region between $x$ and $x'$. To be more precise,
we expect there may be some dissipation of the accumulated energy
to nearby coordinates. For simplicity, however, throughout the paper
we ignore all such effects, and focus solely on individual 2-point
functions.

Using this more complex spacetime dynamics induced by the inhomogeneous
vacuum density, it was proposed that the equation of motion of $\nabla H$
goes as

\begin{equation}
\ddot{a}+\Omega^{2}(t,x)a=0
\end{equation}
where
\begin{align}
\Omega^{2}(t,x) & =\frac{4\pi G}{3}\left(\rho+\sum_{i=1}^{3}P_{i}\right),\quad\rho=T_{00},\quad P_{i}=\frac{T_{ii}}{a^{2}}
\end{align}
In this case, (7) simply states that $\nabla H$ has the behavior
of a harmonic oscillator. That is, it constantly fluctuates around
its equilibrium point. Of course, every crossing of the equilibrium
point is associated with a change of sign.

Let us briefly summarize the spacetime back-reaction to the conjectured
in {[}1{]} constantly fluctuating inhomogeneous vacuum density. Physically,
(5,7) describe a picture where the separation between a pair of neighboring
points, $\Delta x\equiv\left|x-x'\right|$, is constantly fluctuating
between phases of expansion and contraction. In fact, when $\Delta x$
(herein defined as the proper distance) is expanding in some region,
a neighboring region must be contracting, and vice versa. When a vacuum
fluctuation with local energy density \emph{above} its equilibrium
point is considered between $x$ and $x'$, $\Delta x$ (i.e. the
proper distance) grows. On the other hand, if the local energy density
is \emph{below} the equilibrium point, the proper distance $\Delta x$
decreases. This enhanced spacetime dynamics is characterized by constant
local phase transitions between expansion and contraction which happen
as the local energy density of the vacuum goes through its equilibrium
point, and changes sign as it does so. Those constant phase changes
accumulate, and lead to massive cancellations on cosmological scales.
However, assuming a slight \emph{positive} excess, we get the observed
cosmological expansion. Although the microscopic values of $\rho$
may be huge, their infrared effects are small, \emph{i.e.} the wild
fluctuations do not lead to $\mathcal{O}(1)$ corrections in weak
gravitational regimes. Besides, (8) tell us that the scale factor
$a(t,x)$, in a given neighborhood, depends on the time-dependent
frequency $\varOmega(t,x)$ which exhibits quasiperiodic dynamics.

\subsection*{2.2. Unitary black hole evolution and horizon fluctuations via quantum
vacuum fluctuations}

In {[}2{]} we argued that treating gravity in the near-horizon region
of Schwarzschild black hole as a field theory, and considering its
coupling to the matter fields in this region $\int\phi^{\mu\nu}T_{\mu\nu}$
produces fluctuations which modify the general relativistic description.
We considered fluctuations with local energy density \emph{below,}
and \emph{above} $\zeta$, where $\zeta$ is an arbitrary threshold.
We then studied how the black hole geometry back-reacts to those fluctuations
in two distinct regions, \emph{i.e.} \emph{onto} the horizon, and
just outside the black hole. 

More precisely, we studied how the horizon geometry back-reacts to
\textquotedbl{}strong\textquotedbl{} quantum fluctuations, and how
the near-horizon region back-reacts to \textquotedbl{}weak\textquotedbl{}
fluctuations. The analysis was carried out under the assumptions that
(i) black holes are fast scramblers {[}11{]}, (ii) quantum information
is found in the emitted Hawking particles, and (iii) the scrambled
infallen information \emph{need not} be embedded uniformly across
the horizon.

Let us now precisely define what we mean by \textquotedbl{}strong\textquotedbl{}
and \textquotedbl{}weak\textquotedbl{} quantum fluctuations. 

(A) In a broader sense, we take a fluctuation to be \emph{strong}
if, when considered at asymptotic infinity, its local energy density
leads to a localized particle production

\begin{equation}
a_{i}^{\dagger}\left|0\right\rangle =\left|x\right\rangle 
\end{equation}
Therefore, if we had a measuring apparatus counting the strong fluctuations
in a given spacetime region $\Sigma$ at asymptotic infinity, its
results would be consistent with the expectation value of the number
operator $\left\langle N\right\rangle $ in that region

\begin{equation}
\left\langle N\right\rangle =\sum_{i}^{\mathscr{N}}\int_{\sum}\varphi_{i}^{strong}
\end{equation}
More specifically, in a black hole background, we argue a \textquotedbl{}strong\textquotedbl{}
quantum fluctuation \emph{onto} the horizon yields brief departure
from local quantum field theory. 

(B) A quantum fluctuation is taken as \emph{weak} if its local energy
density is below the threshold $\zeta$. Because of the small local
energy density, we assume the back-reaction of the background metric
would be negligible if we considered weak fluctuations in a relatively
small part of the near-horizon region. That is why we are interested
in how the near-horizon metric back-reacts when the weak fluctuations
are taken on scale $\mathcal{O}(r_{S})$. 

We argue that weak fluctuations on scale $\mathcal{O}(r_{S})$ lead
to non-perturbative effects which manifest in potentially observable
metric fluctuations that can play an important role in observer complementarity
and in gravitational-wave astronomy in the form of detectable \textquotedbl{}echoes\textquotedbl{}
and deviations from general relativity close to the horizon. We assume
that away from a black hole the weak fluctuations do not lead to perturbations,
and leave the geodesic equation invariant. One should keep in mind
that, although effectively negligible at infinity, when examined locally,
the weak fluctuations still cause a pair of points to rapidly change
phases between expansion and contraction.

To demonstrate how we define the weak fluctuations in the vicinity
of the horizon we adopted the following analogy. Suppose we interpret
individual fluctuations as harmonic oscillators, denoted by $\chi_{i}$.
Imagine we place the harmonic oscillators on a string in the vicinity
of the horizon, Fig. 1. Using an arbitrary normalized spacing $\epsilon$,
we can generally describe the string as

\begin{equation}
\sum_{i=1}^{N}\int_{S2}d\varphi(n_{i}\varphi_{i})
\end{equation}
where $\varphi_{i}$ is the oscillation frequency of the different
harmonic oscillators.

In this picture we get an ensemble of fluctuations which, as we will
argue later, yield coherent Schwarzschild-scale metric fluctuations.
We assume separate harmonic oscillators \emph{need not} have the same
frequency. In fact, due to their limited domain of dependence (of
order the wavelength of the fluctuation), even neighboring harmonic
oscillators have different frequencies. Thus an evolution equation
for a particular spacetime region can be given in terms of a linear
combination of the harmonic oscillators in that region. Like we mentioned
earlier, despite the arbitrarily high (diverging) oscillating frequency
a single harmonic oscillator may have microscopically, its effect
in an infrared cut-off is negligible, and we do not expect $\mathcal{O}(1)$
corrections to the background metric. 

Our analysis in {[}2{]} lead us to the following two main results.

First, we found that \textquotedbl{}strong\textquotedbl{} fluctuations,
considered onto the horizon, lead to nonlocal release of quantum information
via substantial deviations from classicality (for similar results
{[}4,5,6{]}). We also demonstrated the model achieves Page-like evaporation
spectrum {[}7,8{]}, and predicts black holes begin emitting Hawking
particles in time, logarithmic in the entropy, \emph{i.e.} $t_{*}\sim\mathcal{O}(r_{S}\log r_{S})$,
where $t_{*}\ll M$$^{3}$.

Second, we found that \textquotedbl{}weak\textquotedbl{} quantum fluctuations,
considered in the vicinity of the horizon, produce metric fluctuations
which, although locally negligible, were shown to accumulate, and
on scale of order the Schwarzschild radius lead to significant modifications
to general relativity. Physically, the horizon periodically shifts
(with frequency $\omega$) radially outward with an amplitude $\delta,$
where $l_{P}\ll\delta<M$. In Schwarzschild coordinates the metric
fluctuations translate to shift of the horizon from $r=2M$ to $r=2M+\delta$.
One can think of the weak fluctuations as exerting a drag-like force
on the black hole which produces observable macroscopic quantum gravity
effects. Schematically
\begin{equation}
\omega\sim M_{BH}^{-1}
\end{equation}
where $\omega$ is the frequency of the metric fluctuations.

The inverse proportionality between the frequency of the metric fluctuations
and the mass of the black hole was argued to lead to thermodynamic
instability at late times. Imagine a freely evaporating Schwarzschild
black hole with no perturbations being introduced to it. Given $M_{BH}=1/T$,
we expect $T$ to monotonically increase as the black hole evaporates.
As we can see, (12) dictates that as the black hole loses mass, $\omega$
steadily increases. Tracing that evolution to the final stages of
the evaporation we assume the black hole becomes thermodynamically
unstable as $M_{BH}\rightarrow m_{P}$, where $m_{P}$ is the Planck
mass. At that point, the black hole was conjectured to explode {[}17{]}.
Here, late-time black hole evolution is in agreement with the thermodynamic
instability first proposed in {[}17{]}. 

Recently, the authors of {[}15{]} numerically solved the Einstein
equations modified with metric fluctuations. Similar to the current
work, they showed that the proposed metric fluctuations lead to major
deviations from the traditional general relativistic black hole description.
In agreement with our results, (in particular (12)), their analysis
shows there is an inverse proportionality between the frequency of
the fluctuations, and black hole's mass. Where for a black hole binary
merger, the overlap between the classical and the modified pictures
of the waveform of the emitted gravitational waves decreases as the
mass of the binary gets smaller. That is, as the frequency of the
metric fluctuations increases, the near-horizon region deviates from
general relativity even more significantly.

\subsection*{3. The quantum vacuum origin of metric fluctuations and nonlocal
evaporation}

In the current Section we adopt the Unruh et al. model, and study
its effects in Schwarzschild background. Consequently, we propose
a dynamical mechanism for a form of strong nonviolent nonlocality
which significantly modifies the traditional field theory picture
in the near-horizon region. Specifically, we demonstrate how the conjectured
in {[}1{]} inhomogeneous vacuum density leads simultaneously to quantum
metric fluctuations {[}2,4,5,6,14,15,16{]}, and nonlocal release of
quantum information. One should note we do not attempt to quantize
gravity in the current paper. Although the considered vacuum fluctuations
may not be normalized, which modifies the stress-energy tensor and
makes it non-generic, they still contribute to its expectation value.

We aim to make the transition from cosmological scales to a black
hole case more consistent. That is why we would like to first expand
more on the implications of (1) for black hole backgrounds.

So let us now focus in more detail on how a pair of neighboring spacetime
points $x$ and $x'$ is affected by the suggested extremely inhomogeneous
vacuum density. Consider the following {[}1{]}

\begin{equation}
\Delta\rho^{2}(\Delta x)=\frac{\left\langle \{T_{00}(t,x)-T_{00}(t,x'))^{2}\}\right\rangle }{\frac{4}{3}\left\langle T_{00}(t,x)\right\rangle ^{2}}
\end{equation}
where $T_{00}(x)$ is the vacuum density at $x$, and $\Delta x\equiv\left|x-x'\right|$
denotes the separation between $x$ and $x'$.

Recall our earlier discussion about the evolution of the local energy
density. Specifically, $\rho\rightarrow0$ as $\Delta x$ gets large,
in which case $T_{00}(x)$ and $T_{00}(x')$ evolve independently. 

Keeping that in mind, suppose a classical Schwarzschild solution

\begin{equation}
ds^{2}=-(1-2M/r)dt^{2}+(1-2M/r)^{-1}dr^{2}+r^{2}d\varOmega^{2}
\end{equation}
where $d\varOmega^{2}=(d\theta^{2}+sin^{2}\theta d\phi^{2})$.

Particularly, we are interested in rewriting (13) in Schwarzschild
geometry. Working in a black hole background requires us to take into
account the large gradient with respect to the horizon
\begin{equation}
\rho=\partial_{r}(\partial_{r}\phi)^{2}
\end{equation}
where $\rho$ is the local energy density, and $\phi$ is the gravitational
potential. Thus near a matter source (15) is very sensitive to changes
in the radial coordinate.

Therefore, by taking into account (15), we can rewrite (13) in Schwarzschild
background as

\begin{equation}
\int_{R}\int_{x}^{x'}J^{2}=\int_{R}\int_{x}^{x'}\frac{\left\langle \{\partial_{r}J(t,x)-\partial_{r}J(t,x'))^{2}\}\right\rangle }{\frac{4}{3}\left\langle \partial_{r}J(t,x)\right\rangle ^{2}}
\end{equation}
where $R$ denotes the near-horizon region, and $J$ is the vacuum
accumulation term.

Notice that since we no longer work on cosmological scales, (15) becomes
relevant, and so we substituted the vacuum energy density terms, defined
on particular spacetime points, with $J$. Both terms $J(x)$ and
$J(x')$ are calculated with respect to the horizon. Here, $J$ is
very sensitive to radial changes because of (15). In case $x$ and
$x'$ both lie on the same $r=const$ surface, the expectation value
of $J$ between them varies stochastically. 

Because $J$ is of particular importance in the near-horizon neighborhood,
let us now focus on its general properties.

Since the local energy density in a spacetime region which includes
a black hole exhibits Gaussian-like distribution, (15), we assume
that as $r\rightarrow\infty$, $\left\langle J\right\rangle \rightarrow0$.
Although this may be true for the expectation value of $J$, its \emph{actual}
value must still constantly fluctuate due to the extremely inhomogeneous
vacuum density. Clearly, we can see that in the near-horizon region,
$J$ has a nonzero expectation value $\int_{R}\left\langle J\right\rangle \gg0$.
Therefore, the strongly radial-dependent behavior of $J$ in the near-horizon
region is trivially given as 

\begin{equation}
\int_{R}\frac{\partial\left\langle J\right\rangle }{\partial r}
\end{equation}
Where depending on $r$ with respect to the horizon, the general solutions
to (17) are

\begin{equation}
\int_{R}\frac{\partial\left\langle J\right\rangle }{\partial r}=\begin{cases}
0\quad\:\:for & r>r_{S}\\
>0\:\:for & 2M<r<r_{S}
\end{cases}
\end{equation}
Lastly, we can easily extend (17) to include an arbitrary number of
gauge fields as

\begin{equation}
\int_{R}\left\langle T_{\mu\nu}\right\rangle +\frac{\partial\lambda}{\partial r}
\end{equation}
where $\lambda$ is the vacuum, and $T_{\mu\nu}$ is the stress-energy
tensor. 

Note that at constant $r$ from the horizon, the energy density of
the vacuum fluctuations depends on the internal degrees of freedom
of the black hole, and varies stochastically.

\subsection*{3.1. Metric fluctuations: weak quantum fluctuations = local phase
transitions}

In this subsection we demonstrate how the proposed inhomogeneous vacuum
density can modify the classical near-horizon physics, and thus yield
metric fluctuations. Specifically, by embracing the harmonic-oscillator-like
constant phase transitions of $\nabla H$, we rewrite (5) in Schwarzschild
background, and study how the metric back-reacts. Hence, we show that
in this scenario one can obtain the conjectured quantum corrections
to general relativity in the region just outside the black hole. Like
we saw earlier, the microscopically extremely inhomogeneous vacuum
density gives freedom to the scale factor to be locally expanding
or contracting. Recall that in (5) we defined a local Hubble rate
term $\nabla H$, and argued that, due to the inhomogeneous vacuum
density, it has the dynamics of a harmonic oscillator (7). Namely,
it constantly changes between phases of expansion and contraction.

\begin{figure}
\includegraphics[scale=0.53]{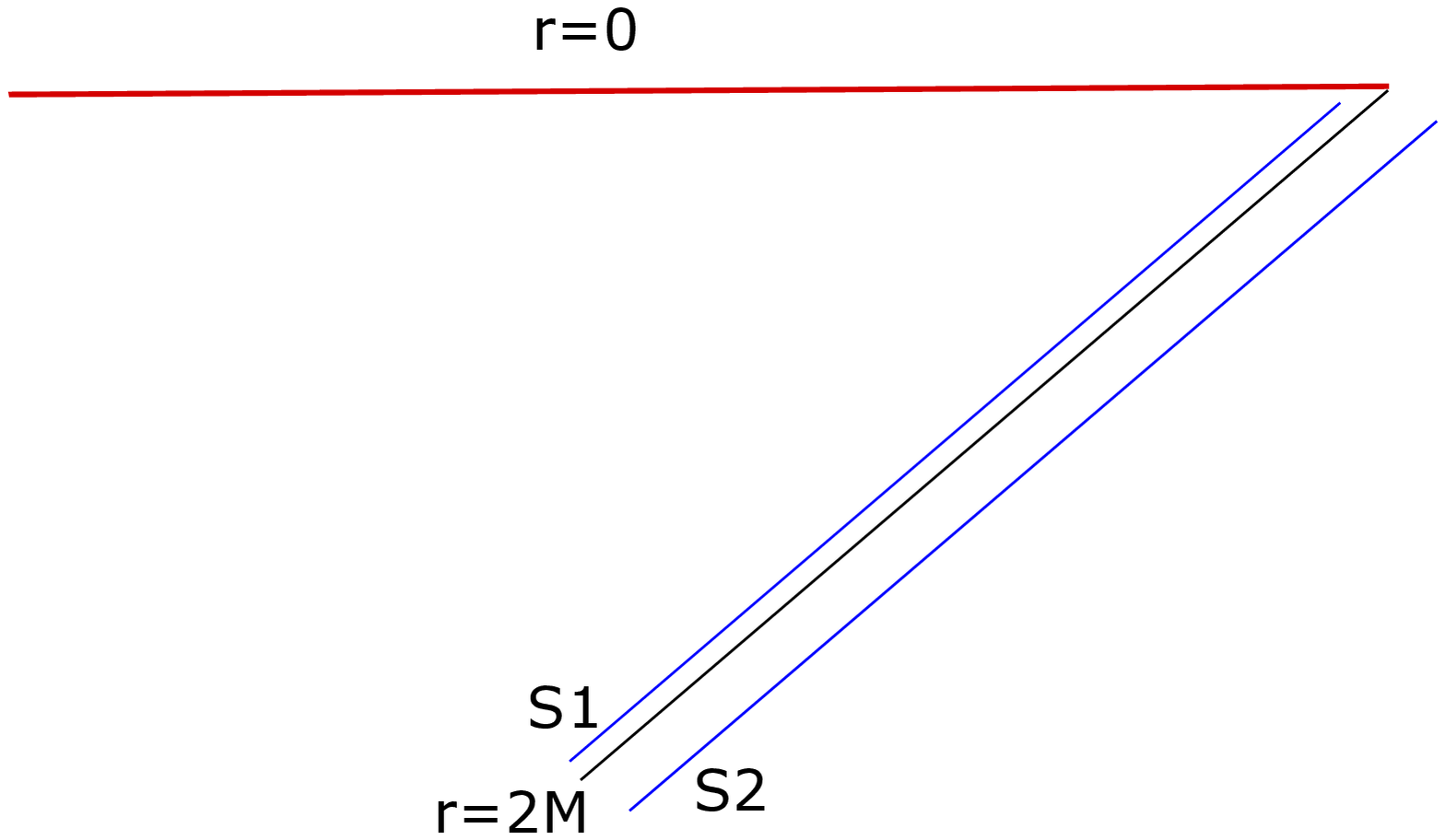}

\caption{The bold red line is the singularity, and the bold black line, $r=2M$,
is the horizon. Imagine that $S1$ coincides with the horizon, and
thus modes located on $S1$ are spacelike separated from an observer
outside the black hole. Further, imagine that $S2$ is placed in the
vicinity of the horizon.}
\end{figure}

For the sake of completeness we will examine coarse-grained, and fine-grained
version of the argument.

\subsubsection*{Coarse-grained}

In the particular case we neglect contributions coming from individual
degrees of freedom, and instead only focus on the effective (macroscopic)
back-reaction on scale of order the Schwarzschild radius.

Keeping in mind the large field-strength radial dependence in this
region (15), we can rewrite (5) as

\begin{equation}
\int_{S1}^{S2}\nabla H=-4\pi G\int_{S1}^{S2}\partial_{t}\left\langle J\right\rangle 
\end{equation}
Unlike earlier, where we considered $\nabla H$ between the neighboring
points $x$ and $x'$, we now evaluate it in the region between \emph{S2}
and \emph{S1} (the horizon), Fig. 1.

To better demonstrate the back-reaction of the near-horizon geometry
(20), consider the following \emph{gedanken} experiment. Imagine \emph{S2}
is a timelike hypersurface the size of the Schwarzschild radius just
outside the black hole, coupled to the horizon degrees of freedom.
Suppose we wish to evaluate the constant $\nabla H$ phase transitions
in the region between the hypersurface and the horizon. We assume
this region constantly undergoes uniform phase transitions with characteristic
oscillation cycle time $T$. Where by \textquotedbl{}uniform\textquotedbl{}
we mean that once every $T$, a phase change of $\nabla H$ in that
region takes place. Unlike the cosmological case {[}1{]}, here we
assume that on every $T$ the phase transitions average out to a small
\emph{negative} value. Here, $T$ is given as 
\begin{equation}
T=2\pi/\varOmega
\end{equation}
 where $\varOmega$ is time-dependent frequency which depends on $\left\langle T_{00}\right\rangle $.

That constant change between phases of expansion and contraction just
outside the black hole (assuming a slight \emph{negative} excess every
$T$) yields the dragging-like effect on the horizon. For simplicity,
one can imagine the conjectured dynamics outside the black hole as
the back-reaction of the horizon to a vector field coupled to matter
fields in Rindler space, where we consider only contributions from
modes very close to the horizon.

Physically, the back-reaction of the near-horizon region manifests
as a slight outward shift of the horizon from $r=2M$ to $r=2M+\delta$
in Schwarzschild coordinates.

Imagine a far away observer who cannot probe the region near the horizon.
As far as she is concerned, she may effectively interpret those metric
fluctuations as a Planck scale structure just outside the horizon
(similar to the stretched horizon in observer complementarity or gravitational
wave \textquotedbl{}echoes\textquotedbl{}). Different proposals have
recently been made regarding the possibility of observing such quantum
gravity effects (see Ref. {[}14-16{]}).

In conclusion, we can see that by rewriting (5) in a Schwarzschild
black hole background we can effectively derive the proposed in {[}2-6{]}
and {[}14-16{]} metric fluctuations. 

\subsubsection*{Fine-grained}

In this case we focus on contributions coming from individual degrees
of freedom, \emph{i.e.} phase transitions of $\nabla H$ between a
pair of corresponding points (say, $x$ and $x'$), where both lie
on the same spacelike hypersurface $S_{i}$ in the near-horizon region.
Namely, $x\in S1$, $x'\in S2$, and $x,x'\in S_{i}$, Fig. 2.
\begin{figure}
\includegraphics[scale=0.3]{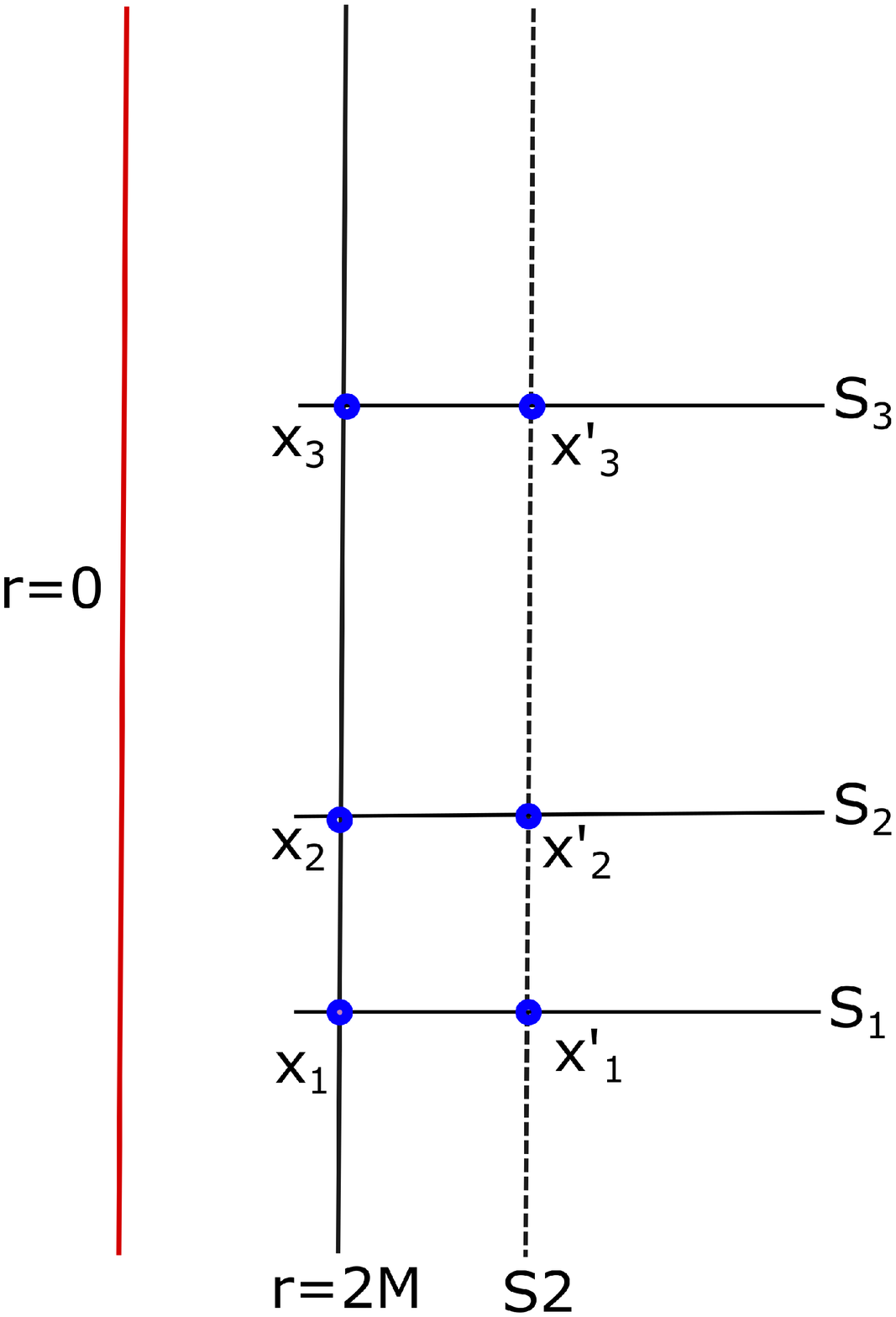}\caption{Similar to Fig. 1, imagine\emph{ $S1$} coincides with the horizon.
The dotted line, \emph{$S2$, }is an artificial (non-physical) timelike
hypersurface in the vicinity of the horizon. Note that the blue dots
$x_{i}$ and $x_{i}'$ are pairs of corresponding points, where $x_{i}\in S1$,
$x_{i}'\in S2$, and $x_{i},x_{i}'\in S_{i}$. Also, $\sum_{i}x_{i}\equiv S1$
and $\sum_{i}x'_{i}\equiv S2$.}

\end{figure}

Suppose we foliate the black hole spacetime into $t=const$ spacelike
hypersurfaces. The particular foliation $\mathcal{F}$ forms a family
of extrinsically flat slicing. The slicing is generic and compatible
with the Killing symmetries of the spherically-symmetric spacetime.
Thus $\mathcal{F}$ preserves the spherical symmetry of the horizon
geometry. The foliation can be intuitively identified with a family
of observers that only have time and radial components. We have chosen
this more trivial geometry-preserving foliation as we aim to make
the study of the back-reaction of the metric to individual 2-point
functions easier.

Considering the particular foliation, we can straightforwardly rewrite
(20) as

\begin{equation}
\sum_{i=1}^{\mathscr{N}}\nabla H_{i}=-4\pi G\sum_{i=1}^{\mathscr{N}}\left\langle J(x_{i},x'_{i})\right\rangle 
\end{equation}
where $x$ and $x'$ lie on the same spacelike hypersurface $S_{i}$,
and also $\sum_{i}x_{i}\equiv S1$ and $\sum_{i}x'_{i}\equiv S2$.
We think of \emph{S2} as an artificial (non-physical) $r=const$ timelike
hypersurface, and we assume it respects the same foliation as the
horizon, Fig. 2.

Evidently from (22), and in agreement with what we have argued earlier
(5), the phase transitions of $\nabla H$ between a pair of neighboring
points depend on the vacuum accumulation term $J$ between them. Note
that due to the stochastic nature of the inhomogeneous vacuum density,
the r.h.s. of (22) fluctuates constantly with field strength of order
the expectation value of the vacuum density in the region $\left\langle J(x,x')\right\rangle \approx\int_{x}^{x'}\left\langle T_{00}\right\rangle $. 

However, (22) does not suffice for a complete description of the near-horizon
dynamics. In addition, we also need to examine the evolution of the
local scale factor. The equation of motion of $a(t,x)$ away from
matter source is given as {[}1{]}

\begin{equation}
tan\varTheta_{x}=\frac{\varOmega(0,x)}{\varOmega(0,x')}tan\varTheta_{x}+\frac{4\pi G}{\varOmega(0,x')}\int_{x}^{x'}J(0,x')dl'
\end{equation}
where $\varTheta_{x}$ is the initial phase of $a(t,x)$ at some arbitrary
$x$.

Similar to what we did with (22), we now wish to express the equation
of motion of $a(t,x)$ in the vicinity of the horizon. We do that
in terms of the vacuum accumulation between a pair of corresponding
points both of which lie on the same spacelike hypersurface. Neglecting
the initial phase of $a(t,x)$ because of its constant fluctuations,
we get

\begin{equation}
tan\varTheta_{x,x'}=\int_{x}^{x'}\frac{\varOmega(x)}{\varOmega(x')}tan\varTheta_{x,x'}+\frac{4\pi G}{\varOmega(x')}\left\langle J(x,x')\right\rangle 
\end{equation}
where $\left\langle J(x,x')\right\rangle \approx\int_{x}^{x'}\left\langle T_{00}\right\rangle $.
One should recall we focus on the small \emph{negative} excess of
the fluctuations every $T$.

The back-reaction of the near black hole geometry to the fluctuations
between a pair of points on a single slice, (22), is negligible, regardless
of their microscopically singular expectation value. However, the
spatial separation between the pair, $\Delta x$, is constantly fluctuating
in a harmonic-oscillator-like manner between phases of expansion and
contraction due to the inhomogeneous vacuum density. Consequently,
by considering the constant fluctuations of $\Delta x$ on all hypersurfaces\emph{
}of order the Schwarzschild radius, and assuming a subtle \emph{negative}
excess on each slice every $T=2\pi/\varOmega$, the back-reaction
accumulates, and leads to deviations from general relativity which
manifest in metric fluctuations. 

In case $M_{BH}=const$, then $\left\langle J(x,x')\right\rangle $
depends on the dimensionality of the internal Hilbert space of the
black hole, and thus fluctuates stochastically

\begin{equation}
\left\langle J(x,x')\right\rangle \sim\int_{x}^{x'}\left\langle T_{00}\right\rangle \sim dim(\mathcal{H}_{int})
\end{equation}
Therefore, following (15), and considering the microscopic back-reaction
of the metric to the constantly fluctuating inhomogeneous vacuum density,
the action for the near-horizon region reads

\begin{equation}
I_{near}=\int_{R}\int dx\int dt\left[\partial_{r}\lambda+T_{00}(x,x')\partial_{r}(x,x')\partial_{t}\lambda\right]
\end{equation}
where the radial-dependent term of the vacuum $\partial_{r}\lambda$
is given by (15). The time-dependent vacuum term $\partial_{t}\lambda$
is taken on a $r=const$ surface with respect to the horizon, and
varies with the change in the mass of the black hole. The action is
given in terms of couplings between the inhomogeneous constantly fluctuating
and radially-dependent vacuum term $\lambda$ and the local energy
density.

In summary, we demonstrated that when we consider the accumulated
back-reaction of the background metric to the weak quantum vacuum
fluctuations (\emph{i.e.} constant phase transitions of $\nabla H$
on different hypersurfaces) across the whole horizon area, $\mathcal{O}(r_{S})$,
we effectively get the proposed quantum metric fluctuations. The current
derivation of these \textquotedbl{}soft\textquotedbl{} quantum modifications
to the effective black hole geometry may be thought of as the microscopic
origin of the metric fluctuations proposed by Giddings {[}4-6{]}.

\subsection*{3.2. Nonlocal black hole evaporation via strong quantum vacuum fluctuations}

In this subsection we continue the study of the back-reaction of a
Schwarzschild black hole to the constantly fluctuating inhomogeneous
vacuum density. More precisely, we examine what effects the conjectured
in {[}2{]} strong quantum vacuum fluctuations have, when they are
considered \emph{onto} the horizon. Specifically, we focus on how
the locality constraint of local quantum field theory is modified
due to the back-reaction of the metric to those fluctuations. As a
result, we argue the horizon geometry \emph{need not} always respect
locality.

Given the fundamental degrees of freedom are not continuously distributed,
we assume the semiclassical geometry to be just an effective field
theory. Thus we expect the local quantum field theory evolution to
only be approximately correct. As a result, deviations from it should
be present in high energy regimes. For instance, black holes are one
such place where we expect nonlocal corrections to manifest. That
is the case since due to the large field-strength radial dependence
in the near-horizon region, $\left\langle T_{00}(x)\right\rangle \gg\left\langle T_{00}(y)\right\rangle $,
where $2M\leq x\leq R$ and $y>R$. 

The particular strong vacuum fluctuations are assumed to have high
enough local energy density as to yield \textquotedbl{}soft\textquotedbl{}
(\emph{i.e.} brief and highly localized) nonlocal corrections to local
quantum field theory. Similar to {[}1{]}, we assume these soft corrections
manifest in the form of local singularity points. 

Because in {[}1{]} the local scale factor $a(t,x)$, defined at each
spacetime point, was argued to have a harmonic-oscillator-like behavior,
as it changes phases, $a(t,x)$ must inevitably go through zero, \emph{i.e.}
yield a local singularity point. However, this was shown to not be
physically problematic. Since we interpret the the local scale factor
as a harmonic oscillator, $a(t,x)=0$ is just a generic part of the
oscillation cycle which takes place every $T$. Namely, a harmonic
oscillator cannot change sign (phase) without passing through zero.
Therefore, in a generic oscillation cycle a singularity point disappears
almost immediately.

Similarly here, because we interpret individual strong vacuum fluctuations
\emph{onto} the horizon as harmonic oscillators, the microscopically
singular expectation value (as predicted by quantum field theory)
of a single strong fluctuation at a given point on the horizon should
not be problematic. Likewise, we assume the local singularity point
disappears almost immediately. 

Physically, a strong vacuum fluctuation on the horizon briefly violates
the generic locality constraint of local quantum field theory in that
small region. As a result, this deviation from classicality allows
quantum information to escape to asymptotic infinity at the necessary
rate. Due to the small domain of dependence and brief lifespan of
the local singularity points, their effects on a freely falling observer
are negligible. These nonlocal correction to the semiclassical geometry,
although \textquotedbl{}soft,\textquotedbl{} have significant effect
on the black hole over periods compared to it's lifetime. That being
said, we assume that, since the inhomogeneous vacuum density is constantly
fluctuating, it will carry out quantum information at a rate of $\mathcal{O}(1)$
per light-crossing time as to restore unitary quantum mechanics. Such
local quantum field theory modifications allow evaporation of information-carrying
Hawking particles to begin as early as the scrambling time.

Let's clarify what we mean when we characterize the nonlocal corrections
as \textquotedbl{}soft.\textquotedbl{} The local quantum field theory
deviations (\emph{i.e.} local singularity points) are \textquotedbl{}brief\textquotedbl{}
in the sense that they have very short lifespan of order the lifetime
of the fluctuation. Furthermore, the deviations are considered to
be \textquotedbl{}highly localized\textquotedbl{} (\emph{i.e.} short
wavelength/high momentum) since they manifest on scales of order the
wavelength of the fluctuation. Thus a strong vacuum fluctuation has
a limited domain of dependence, and cannot lead to $\mathcal{O}(1)$
corrections to the background metric of a solar mass black hole. 

Let's now point out an important distinction between the local singularity
points on cosmological scales (away from a black hole) {[}1{]} and
onto the horizon. We claim there is an intrinsic difference in the
stage of the oscillation cycle during which a singularity point is
produced. More precisely, a strong vacuum fluctuation (onto the horizon)
produces a local singularity when, during an oscillation cycle $T$,
it reaches the maximum of its local energy density, which also happens
to be above a certain threshold $\zeta$. On the other hand, the local
scale factor $a(t,x)$ (away from a black hole) produces a singularity
point when it reaches zero during its oscillation cycle.

Moreover, the local singularity points, although similar, should not
be mistaken with spacetime defects (see Refs. {[}12,13{]}). For instance,
(i) at a local singularity point/spacetime defect the curvature is
divergent, and (ii) particle passing through (near) a spacetime defect/local
singularity will be scattered off; \emph{i.e.} experience a local
Lorentz boost. 

\subsection*{4. Casimir stress interpretation}

In {[}1{]} the Casimir effect was used as a tool for illustrating
the effects of vacuum fluctuations. In the current Section we present
a toy model in which we restate the conjectured quantum metric fluctuations
in terms of the Casimir effect in the near-horizon region. Specifically,
we present the metric fluctuations in the language of the Casimir
effect just outside the black hole (between \emph{S1} and \emph{S2},
Fig. 1).

Likewise, we will examine two distinct cases: coarse-grained, and
fine-grained.

In its general form, the Casimir stress equation is given as {[}1{]}

\begin{equation}
S(t,x,y)=T_{zz}^{inside}-T_{zz}^{outside}
\end{equation}
In this picture, imagine the role of the pair of conducting plates
is played by \emph{S1} and \emph{S2}, Figure 1. 

\subsection*{4.1. Coarse-grained}

Similar to Section 3, we are only interested in the effective (macroscopic)
dynamics, and thus neglect contributions from individual degrees of
freedom.

We can straightforwardly expand (27) in the near-horizon region as 

\begin{equation}
\int_{S1}^{S2}S(t,x,y)=\int_{S1}^{S2}\left\langle \partial_{r}\rho_{in}\right\rangle -\int_{S2}^{\infty}\left\langle \partial_{r}\rho_{out}\right\rangle 
\end{equation}
where, as we showed earlier, the expectation value of the radially-dependent
local energy density $\left\langle \partial_{r}\rho_{in}\right\rangle $
is of order the vacuum accumulation term in that region

\begin{equation}
\int_{S1}^{S2}\left\langle \partial_{r}\rho_{in}\right\rangle \sim\int_{S1}^{S2}\partial_{r}\left\langle J\right\rangle 
\end{equation}
Following (18), we expect $\int_{R}\left\langle J\right\rangle \gg0$
and $\int_{\infty}\left\langle J\right\rangle =0$. That is, the second
term on the r.h.s. of (28) vanishes at the asymptotic limit; namely
$\left\langle \rho_{out}\right\rangle \rightarrow0$ as $r\rightarrow\infty$.

Therefore, we generally get

\begin{equation}
\left\langle T_{zz}^{inside}\right\rangle \gg\left\langle T_{zz}^{outside}\right\rangle 
\end{equation}
One should keep in mind that, regardless of whether $\left\langle J\right\rangle $
vanishes or not, it constantly fluctuates due to the inhomogeneous
vacuum density. 

Notice that (30) is in agreement with the general construction of
the Casimir effect, (27).

Considering the microscopic back-reaction of the spherically-symmetric
background metric that we discussed earlier, and the greater energy
density in the vicinity of the horizon (30), one can easily see how
the proposed quantum metric fluctuations of order the Schwarzschild
radius emerge in this setting.

\subsection*{4.2. Fine-grained }

Similar to our approach in Sec. 3(A), we begin by foliating the horizon
region into individual spacelike hypersurfaces. As a result, we can
now express the l.h.s. of (28) as

\begin{equation}
\int_{R}S(t,x,y)=\sum_{i=1}^{\mathscr{N}}\left\langle J(x_{i},x'_{i})\right\rangle 
\end{equation}
Evidently, the non-vanishing l.h.s. of (28) can straightforwardly
be rewritten as a linear combination of vacuum accumulation terms,
defined on individual slices.

Due to the inhomogeneous vacuum density, $\Delta x$ on any given
slice constantly fluctuates between phases of expansion and contraction.
Moreover, because of the strong field gradient, (15), and considering
(30), we assume that in the large $\mathscr{N}$ limit (\emph{i.e.}
of order the Schwarzschild radius) the accumulated back-reaction is
positive. Thus, an observer at asymptotic infinity sees oscillations
of the horizon as a back-reaction of the underlying black hole metric.

In summary, we see that with minimal assumptions one can restate our
earlier argument about quantum metric fluctuations in terms of the
Casimir effect in a near black hole region. 

\subsection*{5. Conclusions}

In the current work we studied how a Schwarzschild black hole back-reacts
to the constantly fluctuating inhomogeneous vacuum density proposed
in {[}1{]}. More precisely, embracing the microscopically singular
expectation value of the local energy density of the inhomogeneous
vacuum fluctuations (predicted by quantum field theory), we examined
how a black hole metric back-reacts in two distinct regions: the vicinity
of the black hole, and onto the horizon. As a result, we demonstrated
that vacuum fluctuations above a given threshold, considered onto
the horizon, cause deviations from local quantum field theory. Meanwhile,
fluctuations below that threshold, considered in the vicinity of the
horizon, lead to potentially observable metric fluctuations of order
the Schwarzschild radius.

Physically, the conjectured modifications of local quantum field theory,
induced by the strong vacuum fluctuations onto the horizon, were argued
to lead to nonlocal release of information-carrying Hawking particles.
In fact, we argued that in this scenario a black hole can begin radiating
quantum information to infinity as early as the scrambling time {[}2{]}. 

On the other hand, we argued that weak fluctuations in the near-horizon
region yield observable macroscopic quantum gravity effects in the
form of metric fluctuations of order the Schwarzschild radius. That
is, constant oscillations of the horizon between $r=2M$ and $r=2M+\delta$.
As far as a distant observer is concerned, we assume she may interpret
the conjectured metric fluctuations as a physical membrane just outside
the horizon. Thus the proposed metric fluctuations may serve as the
microscopic origin of the stretched horizon in observer complementarity
{[}9{]}. Also, we assume the proposed metric fluctuations play a significant
role in binary black hole mergers. In particular, they may produce
observable post merger gravitational wave \textquotedbl{}echoes\textquotedbl{}
similar to {[}18{]}.

The recent advances in gravitational wave astronomy have opened new
possibilities for experimentally testing models, similar to this one,
which predict deviations from general relativity in the near-horizon
region. In addition, the current scenario could also be approached
from an accretion disk perspective as we believe the metric fluctuations
may have measurable effects on accretion disk flows around a black
hole.
\begin{acknowledgments}
This research is supported in part by project RD-08-112/2018 of Shumen
University.
\end{acknowledgments}

\end{document}